\documentclass[iop]{emulateapj}
\usepackage{apjfonts}
\usepackage{xcolor}
\definecolor{darkmidnightblue}{rgb}{0.0, 0.2, 0.4}
\definecolor{midnightblue}{rgb}{0.1, 0.1, 0.44}
\usepackage[backref,breaklinks,colorlinks,citecolor=darkmidnightblue]{hyperref}
\usepackage{apjfonts}
\usepackage{graphicx}
\usepackage{color}
\DeclareGraphicsExtensions{.eps,.png,.jpg}
\usepackage{amsmath}

\input psfig.tex

\newdimen\digitwidth      
\setbox1=\hbox{0}         
\digitwidth=\wd1          
\catcode`"=\active        

\def"{\kern\digitwidth}

\def\apj{{ApJ}}
\def\apjs{{ApJS}}

\def\lax{{$\mathrel{\hbox{\rlap{\hbox{\lower4pt\hbox{$\sim$}}}\hbox{$<$}}}$}}
\def\gax{{$\mathrel{\hbox{\rlap{\hbox{\lower4pt\hbox{$\sim$}}}\hbox{$>$}}}$}}
\def\simlt{\lower.5ex\hbox{$\; \buildrel < \over \sim \;$}}
\def\simgt{\lower.5ex\hbox{$\; \buildrel > \over \sim \;$}}

\def\mnras{{MNRAS}}

\journalinfo{The Astrophysical Journal, 833:2 (11pp), 2016 December 10}
\submitted{Received 2016 February 2; revised 2016 August 16; accepted 2016 September 2; published 2016 December 1}

\begin{document}


\title{Hierarchical Galaxy Growth and Scatter in the stellar mass - halo mass Relation}

\author{Meng Gu\altaffilmark{1}, Charlie Conroy\altaffilmark{1} and 
Peter Behroozi\altaffilmark{2, 3, 4}}                                     
\altaffiltext{1}{Department of Astronomy, Harvard University, Cambridge, MA 02138, USA}
\altaffiltext{2}{Space Telescope Science Institute, Baltimore, MD 21218, USA}
\altaffiltext{3}{Department of Astronomy, University of California at Berkeley, Berkeley, CA 94720, USA}
\altaffiltext{4}{Hubble Fellow}

\begin{abstract}

The relation between galaxies and dark matter halos reflects the combined effects 
of many distinct physical processes.  Observations indicate that the $z=0$ stellar 
mass--halo mass (SMHM) relation has remarkably small scatter in stellar mass at 
fixed halo mass ($\lesssim$ 0.2 dex) with little dependence on halo mass.  We 
investigate the origins of this scatter by combining $N$-body simulations with observational constraints on the SMHM relation.  We find that at the group 
and cluster scale ($M_{\rm vir}>10^{14}{M_\odot}$) the scatter due purely to hierarchical 
assembly is $\approx0.16$~dex, which is comparable to recent direct observational 
estimates.  At lower masses, mass buildup since $z\approx2$ is driven largely by in-situ 
growth.  We include a model for the in-situ buildup of stellar mass and find that an intrinsic 
scatter in this growth channel of $0.2$~dex produces a relation between scatter and halo mass 
that is consistent with observations from $10^{12}{M_\odot}<M_{\rm vir}<10^{14.75}{M_\odot}$.  
The approximately constant scatter across a wide range of halo masses at $z=0$ thus 
appears to be a coincidence, as it is determined largely by in-situ growth at low masses and by 
hierarchical assembly at high masses.
These results indicate that the scatter in the SMHM relation can provide unique insight into 
the regularity of the galaxy formation process.

\end{abstract}


\maketitle


\section{Introduction}
According to the widely accepted cold dark matter model, the assembly history 
of galaxies is largely driven by the hierarchical growth of the underlying dark 
matter structures (e.g.,  \citealp{White:1978uk, Peebles1982, Blumenthal1984, 
White1991}).  In the past several decades, increasingly accurate cosmological 
simulations of dark matter halos have enabled the detailed study of the structure 
formation process and has improved our understanding of the statistics of dark 
matter halos and their evolution (e.g., \citealp{Springel2005, Klypin2011}).

However, due to a multitude of baryonic processes the galaxy formation process 
is much more complicated than the cosmological growth of dark matter halos.  In
addition to the hierarchical assembly history of dark matter structures, star 
formation, feedback from stars and supermassive accreting black holes shape 
the galaxies we see today.  Many of these processes are not yet fully understood.  
The observed scaling relations of galaxies are important to disentangle this 
complex situation and constrain galaxy formation models.  Scaling relations like 
the star-formation ``main sequence''
\citep{Gonzalez2010, Noeske2007a, Noeske2007b}; and the
``fundamental metallicity relation'' (e.g., \citealp{Mannucci2010, Yates2012}),
were discovered in recent years.  So far, many phenomenological models and numerical
simulations can already reproduce some of these relations qualitatively, and in a few cases, 
quantitatively as well. A remarkable aspect of these relations is their small intrinsic scatter 
($\leqslant 0.3$dex) about the the mean trends. 

The information contained in the
scatter of these relations is not fully understood, but should provide additional
insight into the galaxy formation processes \citep{Salmi2012, Whitaker2012, Whitaker2015}.  
For instance, the tight relation between $M_{\ast}$ and star
formation rate (SFR) observed across a wide range of redshifts suggests that the
star-formation within galaxies is a highly regulated process (but see \citealp{Kelson2014} 
for an alternative point of view), and enhanced SFR during
mergers has a small effect on the total mass-buildup of the galaxies \citep{Noeske2007b}.  
The small intrinsic scatter of the fundamental plane of early-type galaxies sheds light 
on the variety of stellar populations of these galaxies \citep{Prugniel1996, Forbes1998, 
Gargiulo2009, Graves2010, Taranu2015}.  Likewise, scatter in the $M_{\ast}-$size relation 
appears to be small and constant since $z \sim 2$
\citep{vanderWel2014}, which provides constraints on simple dry merger models
\citep{Nipoti2012}.  It is not known to what extent the scatter in these relations is
due to different evolutionary pathways or to intrinsically stochastic processes
(\citealp{Kelson2014}). 

By connecting dark matter halos with the properties of galaxies we can understand how
the hierarchical growth of structure regulates the properties of galaxies.  One
powerful tool to link galaxy properties to the underlying dark matter halos is the
stellar mass-halo mass (SMHM) relation.   The mass ratio between the stellar content in
the galaxy and its dark matter halo tells us how efficient the baryonic component has been
converted into stars.  To fully understand the origin of this relation and its intrinsic
scatter, one needs to understand both the hierarchical growth of structures and the
baryonic processes involved.  

\begin{figure*}[t]
\includegraphics[width=18.0cm]{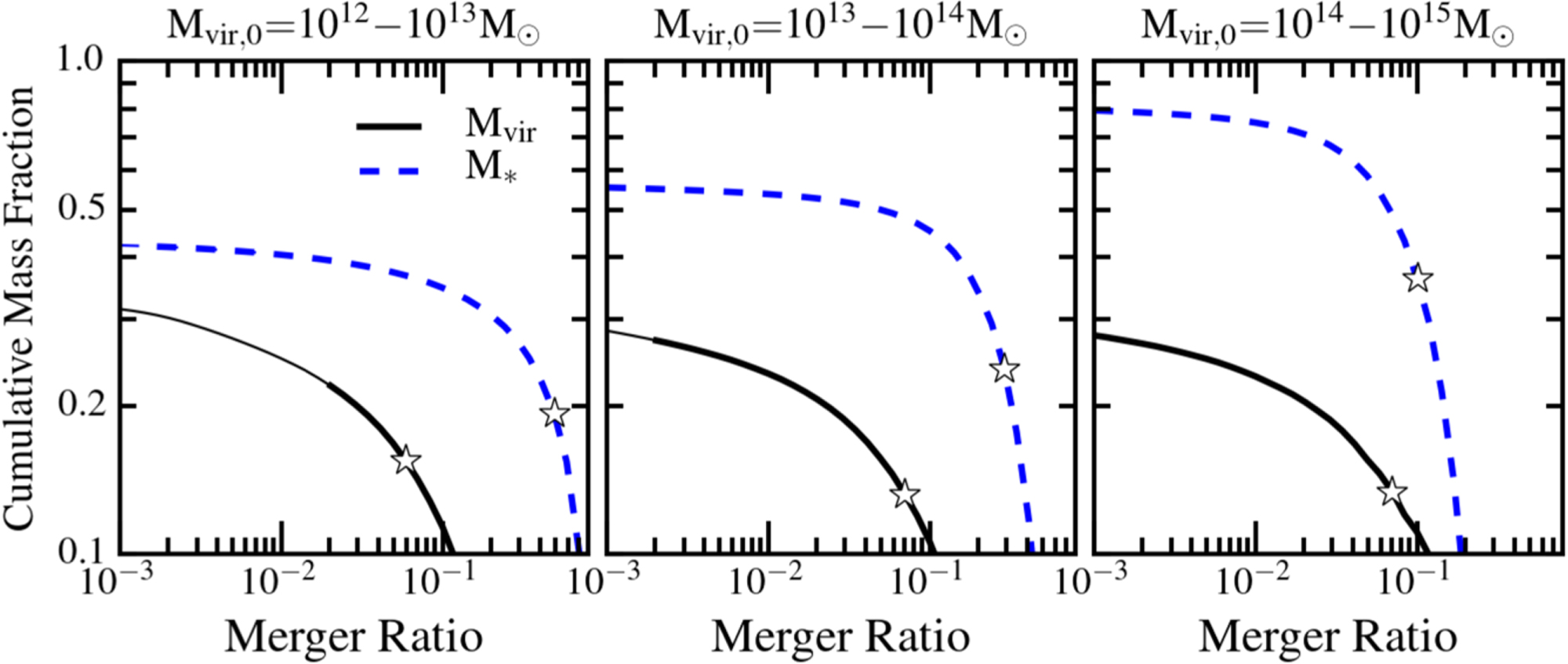}
\caption{
Cumulative stellar and halo mass fraction from accreted objects between $z=2$ and $0$ with 
a merger ratio $>m/M_0$ where $M_0$ is the final mass at $z=0$.  The three panels show the 
cumulative mass contribution from mergers for the local host halos at different scales. 
Blue dashed lines 
and black solid lines show the stellar mass and halo mass fraction, respectively.  
Note that the black lines do not include satellite halos 
that still retain their identity at $z=0$.  Thinner 
lines mark regions below the resolution limit of $2 \times 10^{10} M_{\odot}$ and 
the corresponding stellar mass resolution limit estimated using the SMHM relation from 
\cite{Behroozi2013c}. The dark matter halo merger fractions are expected to be nearly 
self-similar.  The cumulative mass fractions do not converge to one at small mass ratios 
because of the exclusion of satellite halos and smooth accretion 
(for halo mass), and in-situ star formation (for stellar mass). 
Stars mark the merger ratios above which half of the accreted mass is contributed, 
indicating that most of the accreted objects are well resolved by the Bolshoi simulation.  
For example,  the left panel shows that half of the accreted halo mass are through mergers 
with ratio above $0.06$, and half of the stellar mass is accreted through mergers with a ratio above $0.5$.
\label{figure 1}}
\end{figure*}
\noindent

An accurate estimation of both the stellar mass and halo mass is a difficult task.  
Empirical estimates of the SMHM relation are conducted by both direct and indirect 
methods.   Direct measurements include X-ray observations, Sunyaev-Zel'dovich effect, 
galaxy-galaxy lensing and satellite kinematics
within galaxy clusters (e.g., \citealp{Lin2004,Yang2007, Hansen2009, Kravtsov2014}). 
Indirect methods include halo occupation distribution (HOD) modeling
\citep{Leauthaud2011, Leauthaud2012, Zehavi2011, Parejko2013, Guo2014, Yang2003}, 
the conditional luminosity function modeling (\citealp{Yang2009}) and the abundance matching 
technique \citep{Colin1999, Kravtsov1999, Kravtsov2004, Tasitsiomi2004, Vale2004, Shankar2006, Vale2006, 
Neyrinck2004, VanDenBosch2005, Conroy2006, Conroy2009, Behroozi2010, Guo2010, Moster2010, Hearin2013, Reddick2013}.  The scatter in a SMHM relation can be directly estimated from 
samples that have direct dynamical halo mass estimates,  or it can be constrained by galaxy 
clustering statistics, especially at the high mass end where the sensitivity of the 
observations to scatter is large (e.g., \citealp{Zheng2005, Zheng2007, Tinker2007, Leauthaud2012, 
Reddick2013, Zu2016, Shankar2014}).  All of the observational constraints so far suggest that the SMHM relation has a 
remarkably small intrinsic scatter, that shows no evidence for variation with halo 
mass ($\lesssim0.2$ dex scatter in stellar mass 
at fixed halo mass at $z=0$), which is surprising given the complex merging history and the baryonic 
processes involved in galaxy formation.  Simulations and semi-analytic models have 
recently begun to match observational constraints on the SMHM relation 
(e.g., \citealp{Wang2006, Somerville2008, Guedes2011, Munshi2013}).  
There has however been almost no work to 
date on understanding the constraining power contained in the observed \emph{scatter} 
in the SMHM relation.  This is the goal of the present work.

Here we model the evolution of the SMHM relation 
and its scatter induced by the hierarchical growth of dark matter 
halos from $z=2$ to $0$.  We focus on distinct
halos with $M_{\rm vir}>10^{12}{M_\odot}$ at $z=0$.  
During this complex mass assembly history, many processes (e.g., in-situ star formation,
accretion of gas,  stellar/AGN feedback) in addition to mergers affect the evolution of
the SMHM relation.  At the group and cluster scale ($M_{\rm vir}>10^{14}{M_\odot}$), recent 
studies suggest that the bulk of their star formation activity 
has finished by $z\sim2$, and their evolution after $z=2$ is largely 
governed by dry mergers (e.g., \citealp{vanDokkum2001, Treu2005, Thomas2005, 
Choi2014, McDermid2015}).  For lower mass halos, in-situ star formation 
dominates the stellar mass growth.  We therefore also explore the response 
of the scatter due to in situ mass growth.  Our results indicate 
that the hierarchical growth of the dark matter structures induces a strongly 
mass-dependent scatter in the SMHM relation, and the addition of in-situ growth 
with an intrinsic scatter of $0.2$~dex produces a scatter in the SMHM relation consistent with 
observations.

We assume a flat $\rm \Lambda CDM$ cosmology with $h=0.70$, $\Omega_m=0.27$,
$\Omega_{\Lambda} = 0.73$, $\Omega_b = 0.0469$, $n = 0.95$.  The halo mass in this paper,
denoted by $M_{\rm vir}$, is defined as the mass enclosed within a spherical 
overdensity calculated from \cite{Bryan1998}. At $z=0$ the overdensity is 360 times the 
background density.  The stellar 
mass is denoted by $M_{\ast}$.


\section{Methodology}

\subsection{Bolshoi Simulation}
We make use of the Bolshoi simulation~(Klypin et al. 2011), which is a large,
dissipationless simulation of the evolution of the universe, assuming a flat, $\rm \Lambda
CDM$ cosmology.  The simulation adopted the cosmological parameters ($h=0.70,
\Omega_m=0.270, \Omega_{\Lambda} = 0.730, \Omega_b = 0.0469, n = 0.95, \sigma_8 = 0.82$), 
consistent with the results from $WMAP5$ (\citealp{Hinshaw2009, Komatsu2009, Dunkley2009}) 
and $WMAP7$ (\citealp{Jarosik2011, Komatsu2011}).  It followed $2048^3$ particles in a 
comoving volume of $(250h^{-1})^3$ Mpc$^3$ from $z=80$ to $z=0$.  The evolution of the 
dark matter halos are recorded at 180 snapshots.  The simulation has a mass 
resolution of $1.9 \times 10^8M_{\odot}$ and a force resolution of $1h^{-1}$ kpc.  
Assuming a conservative resolution limit of 100 particles \citep{Behroozi2013a}, the halo mass 
resolution limit is therefore roughly $2 \times 10^{10} M_{\odot}$,
allowing us to study the properties of dark matter halos with high accuracy.    

In this work, we use the dark matter halos identified by the \textsc{ROCKSTAR} halo finder
(\citealp{Behroozi2013b}).  \textsc{ROCKSTAR} is based on adaptive hierarchical refinement of
friends-of-friends groups, using not only the six phase-space dimensions but also the
temporal dimension.  Compared with other algorithms, it shows an advantage 
in probing substructure and 
maximizing the consistency of halo properties across time steps (\citealp{Knebe2011}).
The halo catalogs and merger trees are generated by \textsc{Consistent Trees} described in
\citealp{Behroozi2013d}.  This algorithm verifies the consistency of the halo finder and
improves the completeness of the halo catalogs.  It generates the dynamically consistent
merger trees from the Bolshoi simulation with higher accuracy compared with particle based
merger trees.  We use the term subhalo as the halo identified in 
the virial radius of a larger halo.  A halo is defined as a distinct halo if 
it's not within the virial radius of another halo. 

\begin{figure}[t]
\includegraphics[width=8.5cm]{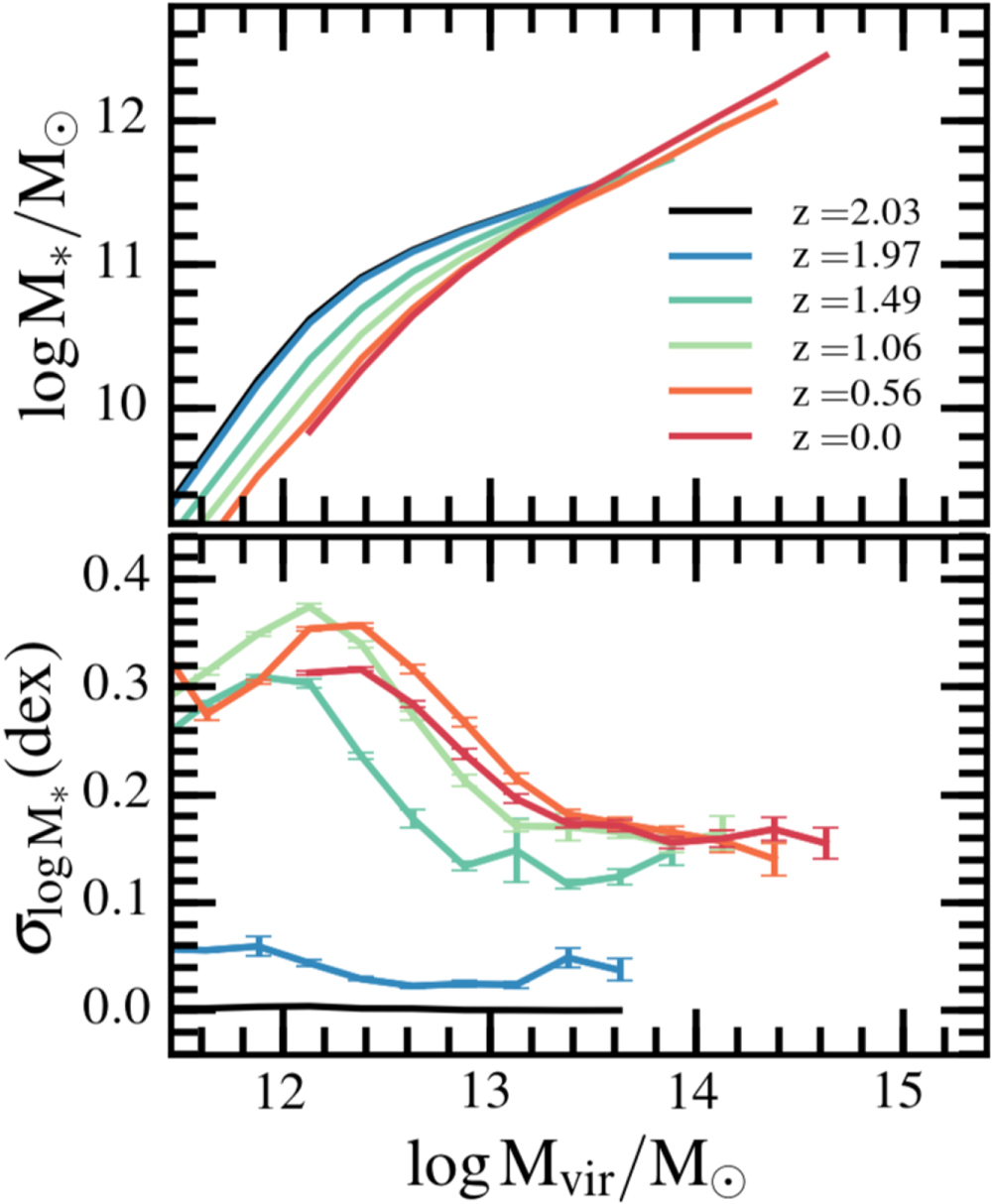}
\caption{
Evolution of the SMHM relation and scatter from $z=2$ to $z=0$ purely 
from hierarchical dark matter halo assembly (i.e., assuming that galaxies do not grow).  
The initial relation at $z=2$ is adopted from \cite{Behroozi2013c}.  
Error bars are determined by bootstrap resampling. 
\label{figure 2}}
\end{figure}
\vskip 0.15cm
\noindent

\subsection{The Stellar Mass-Halo Mass Relation}

Associating galaxies with dark matter halos using the SMHM relation has been done
by a variety of methods, including observations of individual objects, HOD modeling and
abundance matching techniques.  We use the SMHM relation provided by
\cite{Behroozi2013c}.  The authors use a Markov Chain Monte Carlo method to model the SMHM
from $z=8$ to $z=0$ with the aid of a variety of observational data, including
the stellar mass functions, the distribution of specific star formation rates and the
integrated cosmic star formation rates.  The fitting function of the SMHM relations 
has 5 parameters, and can be described by a power law at the low mass end
and a sub-power law at high mass.  The specific form of the fitting function is

\begin{equation*}
	\begin{split}
        \log_{10}\left(M_{\ast}\left(M_{\rm vir}\right)\right)= \log_{10}\left(\epsilon M_1\right)+f\left(\log_{10}\left(\frac{M_{\rm vir}}{M_1}\right)\right)-f\left(0\right) \\
        f\left(x\right)=-\log_{10}\left(10^{\alpha x} +1\right)+\delta \frac{\left(\log_{10}\left(1+\exp\left(x\right)\right)\right)^\gamma }{1+\exp\left(10^{-x}\right)}
	\end{split}
\end{equation*}

\noindent

We adopt the best-fit parameters provided by \cite{Behroozi2013c}. The resulting SMHM 
relation provides a good match to a wide array of observations from $z=8$ to $0$.  
In this work, we consider the
satellite halos that may have a $M_{\rm vir}$ as low as a few times of the halo
completeness limit, as well as the massive host halo more massive than
$10^{14}{M_\odot}$.  We therefore extrapolate the functional form of the SMHM beyond the
observed mass range (\citealp{Behroozi2013c}). The extrapolation at the low-mass end 
does not show a noticeable impact on the mass growth, since there is only very little 
stellar mass content of the extrapolated galaxies.

Besides the stellar content in the galaxies, recent observations and simulations have
demonstrated the important role of the intracluster light (ICL) component to the SMHM at
the high-mass end (e.g.,  \citealp{Lin2004, Gonzalez2007, Gonzalez2013, Zibetti2005}).  
When a satellite galaxy disrupts, it is not clear what fraction of the 
stellar content will be deposited into the central galaxy versus the ICL.  
Meanwhile, it is also not clear whether the ICL can be seen as a separate component, 
and what is its best definition (kinematic or photometric).  Recent observations 
suggest that the formation of the ICL strongly relates to past mergers (e.g., 
\citealp{Gonzalez2013, DeMaio2015}) so that part of the accreted stellar component 
between $z=2$ to $0$ must be in the ICL. In this work, we define the ICL as all the 
stellar components that are accreted onto the host halo but are not bound to 
the central galaxy.  \cite{Behroozi2013c} provides two models of the SMHM
relations across a wide range of redshifts.  One model connects the host halos and the 
central galaxy, and the other uses a stellar component including both the central galaxy and
the ICL.  Throughout the paper, we adopt the SMHM relations including ICL 
as the fiducial model.  We note that we have also explored a model using the SMHM 
without ICL and it does not alter our result of the scatter induced by hierarchical growth.   

\subsection{Modeling Dissipationless Growth} 
To isolate the scatter of stellar mass at fixed halo mass caused by galaxy mergers and
mass accretion, we initially assume that the SMHM relation at $z=2$ has no intrinsic 
scatter.  
(In Section~3.2, we consider a model in which the SMHM relation has a non-zero 
scatter.)
We select a sample of distinct halos with $M_{\rm vir,z=0}>10^{12}{M_\odot}$.  
For each distinct halo at $z=0$, we trace back through its merger tree as a function 
of time up to $z=2$ and find all of its progenitors at different epochs.  The stellar 
masses of these distinct halos at $z=2$ are calculated using its halo mass and the adopted 
zero-scatter SMHM relation at $z=2$.   For each progenitor that was accreted onto 
the host halo between $z=2$ and $z=0$, its stellar mass is estimated in a very 
similar way, using also a zero-scatter SMHM relation at the time of disruption, and the 
peak halo mass ($\rm M_{peak}$) of the progenitor.  The peak halo mass is defined as 
the maximum halo mass in the progenitor's merger history.  During the gradual disruption 
process of a progenitor, its dark matter content is stripped earlier than its baryonic component.  
Hence, the stellar mass of the progenitor should be unaffected in the early phase of the 
merger.  The peak halo mass is therefore a better property to connect
with the stellar component of the progenitor than the total halo mass right 
before disruption (e.g., \citealp{Reddick2013}). 

Figure~1 shows the cumulative mass distribution of both the stellar 
component and the dark matter of the progenitors between $z=2$ and $z=0$ as a 
function of merger ratio.  The halo mass in this figure refers to the mass 
just prior to the accretion. The stellar mass to halo mass ratio is not constant 
as a function of halo mass, therefore the stellar mass contribution from 
progenitors with different halo masses are not proportional.  
Thinner lines in Figure~1 show the regions below approximate halo mass resolution 
limit of 100 particles.  For the two 
highest mass bins more than $90\%$ of the mass is well resolved.  
Even for the lowest mass bin more than $50\%$ of the accreted mass is resolved by the 
Bolshoi simulation.
\begin{figure}[t]
\includegraphics[width=8.5cm]{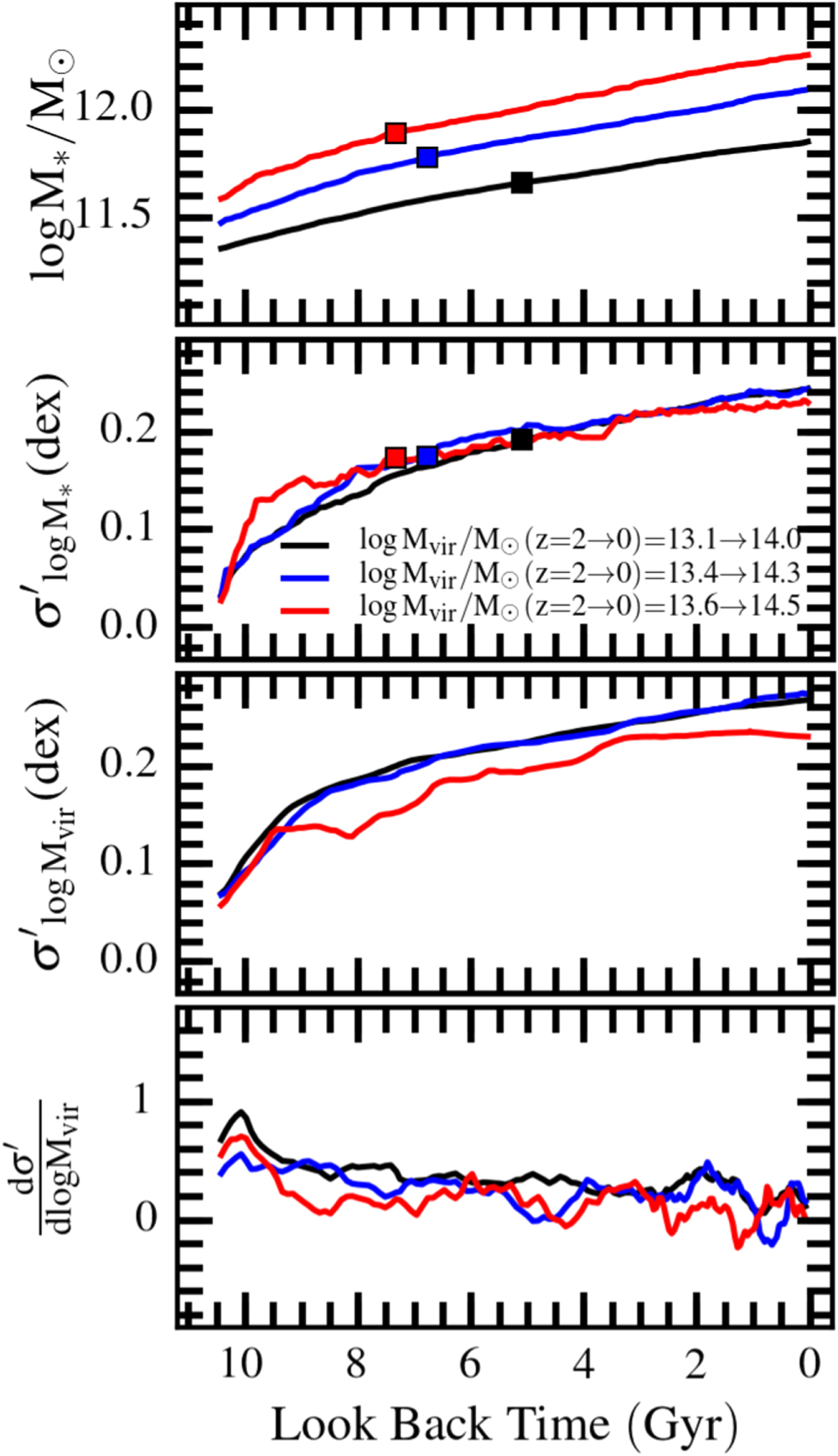}
\caption{
Evolution of the stellar mass and scatter in the three most massive bins from 
$z=2$ to $0$.  From top to bottom, the stellar mass growth, the scatter in 
$\log M_{\ast}$, the scatter in $\log M_{\rm vir}$ and the change in 
scatter per unit $\log M_{\rm vir}$ are shown as a function of look back time.  
Squares indicate the time when the mean stellar mass is doubled.  Note 
that the majority of the growth in the scatter occurs within the first mass doubling time.
\label{figure 4}}
\end{figure}
 
\subsection{Modeling In-situ Growth}

Current models predict that the fraction of the total stellar mass 
contributed by accretion 
depends strongly on galaxy stellar mass.  This fraction can be as 
large as above $80 \%$ for massive 
halos at group or cluster scale, but very small or even negligible for Milky Way-sized 
galaxies (e.g., \citealp{Lee2013, Rodriguez-Gomez2015, Purcell2007, Oser2010}).  
As a result, hierarchical accretion should not be the dominant factor 
influencing the stellar mass scatter at the low mass end.  For this reason, we include 
a model for the in-situ buildup of stellar mass.  We assume that the 
final SMHM relation including both in-situ and ex-situ processes at $z=0$ or $z=1$ should be 
consistent with the result in \cite{Behroozi2013c}.  
To estimate the in-situ stellar mass fraction, in each bin, we adopt a simplified 
assumption and take the average difference between the stellar mass from our hierarchical 
assembly model at $z=0$ and the \cite{Behroozi2013c} model at $z=0$ in this bin as the 
stellar mass growth due to in-situ star formation between $z=2$ and $z=0$.  
The in-situ stellar mass fraction includes both the initial stellar mass estimated 
using the SMHM relation at $z=2$ and the subsequent in-situ growth estimated using the method described above.  
We estimate the in-situ stellar mass fraction at $z=1$ in a similar way.   The resulting $z=0$ in-situ fraction is 
$\sim19\%$ at $M_{\rm vir}=10^{14}{M_\odot}$,  reaches $\sim52\%$ at $M_{\rm vir}=10^{13}{M_\odot}$, and is 
$\sim85\%$ at $M_{\rm vir}=10^{12}{M_\odot}$.  

We also consider a model in which the intrinsic scatter associated with 
the in-situ component has a value of $\sigma_i = 0.2$~dex.  
For the host halos, we assume that the initial stellar masses at $z=2$ are formed 
in-situ and they follow a log-normal distribution with a scatter of $0.2$~dex at fixed halo mass.  
The average in-situ growth after $z=2$ is determined as the average 
difference between the stellar mass from our ex-situ model and the \cite{Behroozi2013c} 
model at $z=0$.  The average total stellar mass due to in-situ growth is the sum of the two.  
At $z=0$, we assume that the total stellar mass due to in-situ growth of 
the host halo follow a log-normal distribution with a scatter of $0.2$~dex at fixed halo mass. 
For the progenitors, we assume that the adopted 
initial SMHM relation is also a consequence of in-situ growth. The SMHM relations 
for progenitors between $z=2$ and $0$ hence also has a $0.2$~dex scatter.

\section{RESULTS} 

We now present our results.  In Section~$3.1$, we show the SMHM relation and its 
scatter evolution driven by purely hierarchical assembly.  We then discuss the 
possible factors that influence the growth of scatter in the case of purely 
dissipationless growth.  In Section~$3.2$ we present results from the model 
including both ex-situ and in-situ processes. We compare our results with observational 
constraints in Section~$3.3$.

\subsection{Scatter due to Ex-situ Growth}

To isolate the influence due to mergers, we first construct a model with only hierarchical 
accretion.  In this section, we present our fiducial ex-situ growth model, and 
then turn to an investigation into the variables affecting the scatter in 
the case of ex-situ growth.  Several sources determine the final scatter in 
$\log M_{\ast}$.  First, the shape of the SMHM relation determines the distribution of 
the stellar mass of satellite galaxies, which in turn influences the 
scatter in $\log M_{\ast}$.  In Section~$3.1.2$, we study 
the relation between the slope of the SMHM relation at $z=2$ and the scatter at $z=0$.  
Second, the growth of dark matter halos due to smooth accretion plays an important role. 
We compare our results with and without smooth accretion in Section~$3.1.3$. 
Finally, 
massive halos on average experience more merger events than low mass halos.  
In Section~$3.1.4$, we show that the scatter of $\log M_{\ast}$ at fixed 
halo mass at $z=0$ also depends on the number of merger events that a distinct halo has
experienced. 

\subsubsection{Fiducial Model}

Figure~2 shows the evolution with time of the SMHM relation and its scatter 
in our fiducial, ex-situ growth-only model.  At $z=2$, the SMHM relation has zero scatter at all 
halo mass.  The stellar mass of the distinct halo gradually builds up through mergers since $z=2$.  
When calculating the SMHM relations, we adopt a bin width of $0.25$~dex in $\log M_{\rm
vir}$.  Only bins that contain more than 30 halos are included in our analysis.  
In order to compute the scatter, we linearly interpolate mean stellar mass 
between bin centers, then subtract off this mean relation before computing scatter.  This avoids 
complications that would arise due to the shape of the SMHM relation and the width of the bins.
The stellar mass and the scatter is the mean and
standard deviation of $\log M_{\ast}$ of galaxies in each bin. 

At $z=1.97$, the first time step right after $z=2$, a 
notable scatter of the SMHM relation is already apparent, especially at the low mass end.  
The scatter then builds up gradually and even shows a slight decrease later on at the high mass 
end.  At $z=0$, the scatter of the SMHM relation is strongly mass-dependent, reaching $0.32$ 
dex at $M_{\rm vir}\sim10^{12.5}{M_\odot}$, which is much larger than the $0.16$ dex scatter 
at the high mass end. 

\begin{figure}[t] 
\includegraphics[width=8.5cm]{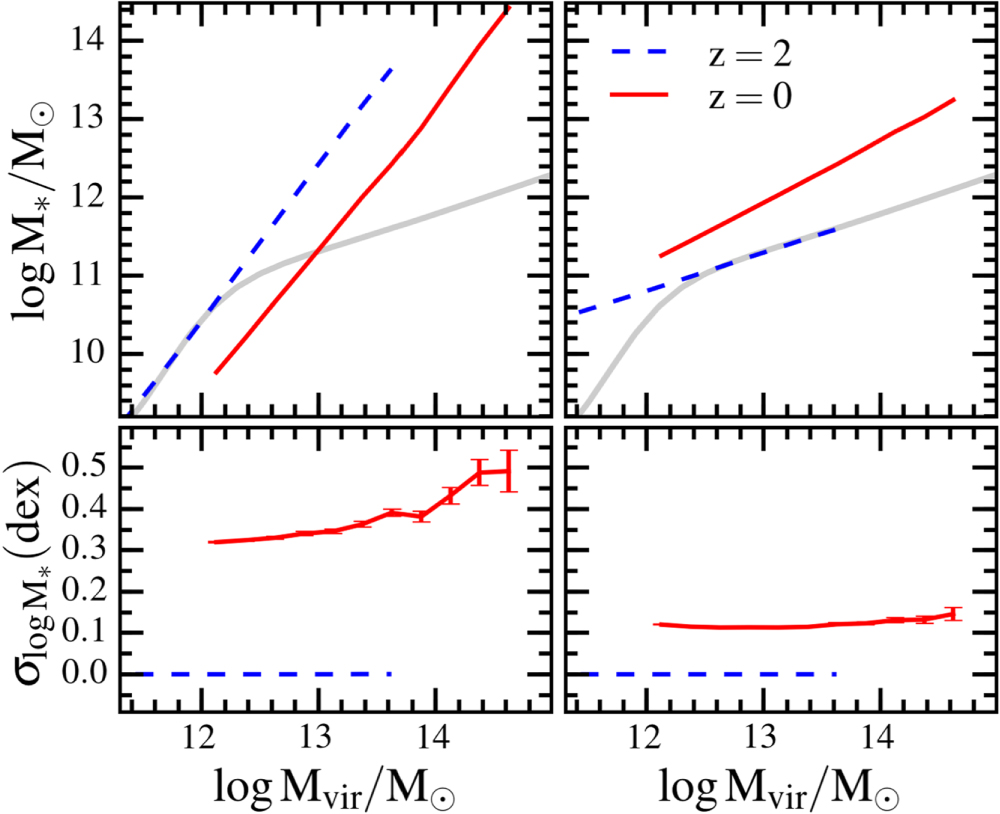}
\caption{
Comparison of the scatter in two hypothetical SMHM relations with single slopes. The 
slopes are identical to the low mass end and high mass end slopes in the SMHM relations 
from \cite{Behroozi2013c}.  Bottom panels show the evolved scatter (red) at $z=0$.  
Note that they are sensitive to the slopes of the SMHM relations.  
A steeper slope at $z=2$ results in a larger scatter of stellar mass at fixed halo mass at $z=0$.
\label{figure 5}}
\end{figure}

\noindent

To investigate the stellar mass growth in a small population of galaxies that share the
same initial halo and hence stellar mass, we identify three groups of halos 
at $z=2$ and follow their evolution to $z=0$.  The result is shown in Figure 3.  
We select the three most massive bins at $z=2$ that
contain more than 30 host halos.  The bins are defined by 
$10^{13.0}{M_\odot}<M_{\rm vir}<10^{13.25}{M_\odot}$, 
$10^{13.25}{M_\odot}<M_{\rm vir}<10^{13.5}{M_\odot}$,
$10^{13.5}{M_\odot}<M_{\rm vir}<10^{13.75}{M_\odot}$ at $z=2$, respectively.  
We trace each halo in each group from $z=2$ to $0$.  Figure 3 shows the growth
in $\sigma^{\prime}_{\log M_{\ast}}$ and $\sigma^{\prime}_{\log M_{\rm vir}}$ as a
function of the look back time.  It is worth mentioning that the $\sigma^{\prime}_{\log
M_{\ast}}$ here is different from the $\sigma_{\log M_{\ast}}$ in Figure~2.  Here we trace
the same host halos in each group from $z=2$ to $0$ and the mean ${\log M_{\rm vir}}$
changes with time, whereas in Figure 2 we study the host halos at fixed halo mass at each
time step.  The squares in Figure 3 show the time when the geometrically averaged 
stellar mass in each group is doubled with respect to $z=2$.  The group with a larger halo mass
double their stellar mass earlier than the low mass groups.  The scatter in all three
groups increases quickly within the time the stellar masses are doubled, and then gradually 
slow down.  The low mass group has a slightly larger final scatter than the high mass 
group.

\subsubsection{The Slope of the SMHM Relation}

In Figure 2, there is an obvious transition in the slope of the SMHM relation at around $M_{\rm
vir}=10^{12.5}{M_\odot}$.  The scatter in $\log M_{\ast}$ shows a dramatic decrease at a 
similar halo mass, suggesting a connection between the slope of the SMHM relation and the
scatter associated with ex-situ growth.  To study the relation between the 
slope of the SMHM relations
for both the distinct and satellite halos and the resulting scatter at $z=0$, we construct 
two hypothetical SMHM relations that roughly describe the high mass end slope and the low
mass end slope of our fiducial SMHM relation.  They are shown in Figure 4. 

The two hypothetical SMHM relations and their scatter at $z=2$ are shown in blue.  
The fiducial SMHM relation is 
plotted as gray lines in the background a comparison. We assume that both the distinct halos 
at $z=2$ and the progenitors between $z=2$ to $0$ follow the same SMHM relation. 
We note that these single power law relations are obviously in conflict with 
the data and are not used to reflect the actual evolution.  We just use them as a 
toy model to show the effect of the slope on the scatter. 
We adopt the same approach described in Section~$3.1.1$ for computing the scatter.  
The SMHM relations and the scatter about the mean relation at $z=0$ are shown in red.  
The slope of the SMHM relations for the initial host halos and the progenitors has 
a clear effect on the scatter at $z=0$.  

Overall a steeper SMHM relation for 
host and satellite halos produces a larger final scatter at $z=0$.  This is consistent with 
our fiducial model shown in Figure~2.  As shown in Figure~1, while dark matter 
halos have a relatively self-similar distribution of progenitors, the accretion of the stellar 
component depends strongly on mass.  This is due to the changing slope in the SMHM relation.  
For a fixed distribution of progenitor halos, a steeper slope implies a wide range of progenitor 
galaxies and hence a larger final scatter.  A second reason is that the slope of the SMHM 
relation influences the relative importance of major vs. minor mergers.  A steeper SMHM 
relation produces fewer major mergers given the same dark matter assembly history, and 
hence increases the Poisson scatter.

\subsubsection{Effect of Smooth Accretion}

Part of the halo mass growth is associated with a diffuse component, i.e. the 
growth of dark matter halos due to smooth accretion.  This contributes about $40\%$ 
to the total growth of the dark matter halo (e.g.~\citealp{Stewart2008, Genel2010}).  
Here we show that the smooth accretion is another 
factor influencing the scatter at $z=0$.

In Figure~5, we compare results with different initial SMHM relations at $z=2$, 
including or not the growth of the dark matter halo due to smooth accretion.  
All of the work up to this point has included smooth accretion of dark matter 
as a component of the total growth.  The smooth accretion is also the major reason that 
the black lines in Figure~1 do not converge to one at small mass ratios.  To estimate the 
scatter without the influence of smooth accretion, we assume that the final halo mass 
of the distinct halo at $z=0$ is the total virial mass of the progenitors at the time just 
prior to the accretion.  The stellar mass is calculated in the same way as described in 
Section~$2.3$, i.e., it's the total stellar mass of progenitors based on their peak halo 
mass, and the SMHM relations at the time of disruption. Therefore the halo mass growth 
of distinct halos is solely due to the halo mergers, and the stellar mass of the central 
galaxy at any time is just the total
stellar mass in the progenitors disrupted before that time.

\begin{figure}[t]
\includegraphics[width=8.5cm]{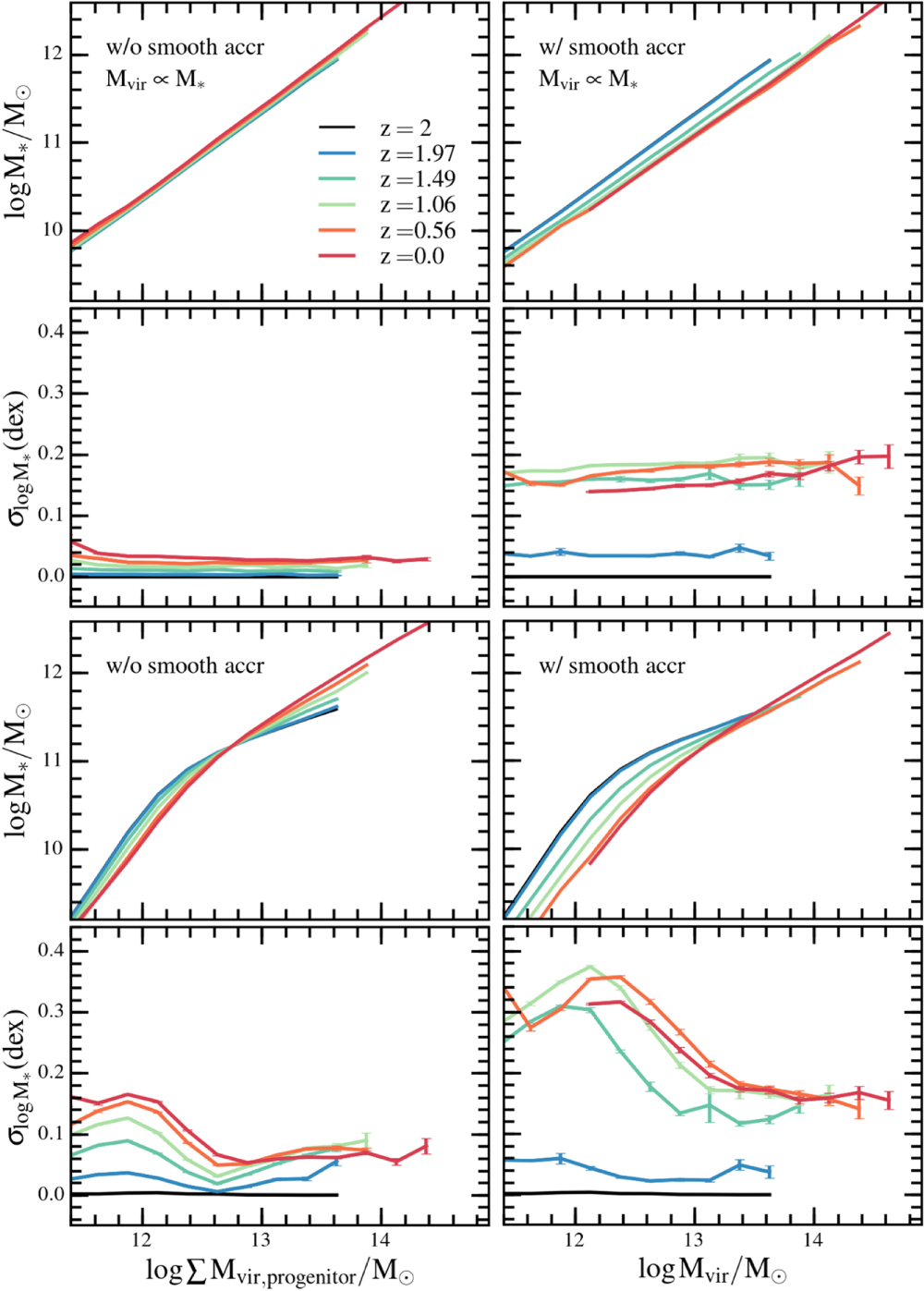}
\caption{
Impact of the shape of the SMHM relation and smooth accretion of dark matter on the scatter.  
Left panels and right panels show the result without and with smooth accretion.  In the 
top panels we adopt a hypothetical SMHM relation in which the halo mass is proportional 
to the stellar mass.  In the bottom panel we adopt the functional
form of the SMHM relation from \cite{Behroozi2013c}.
\label{figure 6}}
\end{figure}
\noindent

In the top two panels we assume a SMHM relation with a fixed ratio between halo 
mass and stellar mass at $z=2$ for both distinct halos and subhalos.  If there 
were no smooth accretion of dark matter, we would predict that the ratio between 
$M_{\rm vir}$ and $M_{\ast}$ should remain almost constant after mergers—except 
for a tiny variation due to mass ejection during major mergers (\citealp{Behroozi2013JCAP, Behroozi2015}).  
The scatter in the SMHM relation is then expected to be nearly $0$.  This scenario 
is shown in the top left panel. It results in an almost uniform and very small
scatter in $\log M_{\ast}$ at fixed halo mass at $z=0$.  We assign the stellar mass based 
on peak halo mass.  The very small scatter in the top left panel comes from the difference 
between the peak mass and virial mass of satellites.  However, when the growth of 
dark matter halos due to smooth accretion is included, the scatter grows 
to about $0.2$ dex from $z=2$ to $z=0$, as shown in the top right panel.  Comparing the two top 
panels, we conclude that part of the final scatter in $\log M_{\ast}$ 
can be attributed to the variation in the final halo mass due to smooth 
accretion.  In other words, even for a model in which the stellar mass did not grow 
but the halos grew via smooth accretion, one would expect the scatter in stellar mass 
\emph{ at fixed halo mass} to increase over time.  

Smooth accretion also has an effect on the scatter for our fiducial SMHM relation.  
This is shown in the bottom two panels of Figure 5, where the smooth accretion induces an 
increase in scatter of about $0.1$ dex in host halos with $M_{\ast}$ 
between $10^{13}{M_\odot}$ and $10^{14}{M_\odot}$, and a much larger increase in scatter at low mass host halos.

\subsubsection{Number of Merger Events}
  
We next consider the effect of the number of mergers on the scatter 
in stellar mass.  To do this we set up a simple Monte Carlo simulation for the 
growth of stellar mass and scatter as a function of the number of merger events.
We compare the
evolution of two groups of halos with different initial virial mass, 
$M_{\rm vir}=10^{12.6}{M_\odot}$ and $M_{\rm vir}=10^{13.6}{M_\odot}$.  Each group contains
500 halos.  We construct a sample of progenitors for each group using the \textsc{ROCKSTAR} halo
catalog of the Bolshoi simulation.  For example, the progenitor catalog of the 
$M_{\rm vir}=10^{12.6}{M_\odot}$ group consists of all the progenitors 
between $z=2$ to $0$ of the host halos with
$M_{\rm vir}=10^{12.6}{M_\odot}$ at $z=2$ in Bolshoi simulation.  The top panel of
Figure~6 shows the mass weighted distribution of the progenitors' stellar mass.  In this
figure, the peak stellar mass of the progenitors is slightly higher for the high mass
group.  We use the number distribution of the progenitors in each group as a probability
distribution and allow one progenitor to merge with the host halo at each step.  
The distributions are quite similar because of the shallow slope of the SMHM 
relation above $M_{\rm vir} \approx 10^{12.5}{M_\odot}$
The bottom
panel of Figure 6 shows the evolution of the geometrically averaged stellar mass in each
group.  

For both groups, the scatter of $\log M_{\ast}$ quickly builds up to $\approx0.15$
dex within the first 300 mergers, reaches a peak scatter near when the geometrically averaged
stellar mass is doubled, and then slowly decreases.  
This behavior can be understood as follows.  Initially the scatter is 
dominated by the in-situ mass, which by construction was initialized to be the 
same within each bin.   Once the ex-situ mass becomes a major component but before 
there have been too many mergers, the scatter is large and influenced by the random 
draws from the progenitor distribution. Eventually, when the number of mergers is very 
large, each galaxy will have fully sampled the progenitor distribution and will 
therefore converge to the same mass in a central limit theorem-like process.
The scatter produced from this simple model is similar with what is seen in 
our fiducial model (Section~$3.1.1$).  This simple simulation reproduces 
the result that the scatter of $\log M_{\ast}$ is slightly smaller for the high 
mass group.  This suggests that the scatter is sensitive to both the 
number of mergers and the mass function of the progenitors.  Although not shown in 
Figure~6, the scatter will slowly decrease to zero
eventually after $\sim10^6$ mergers, when the stellar mass has increased by a factor of $10^3$. 

\vskip 0.2cm
\begin{figure}[t]
\includegraphics[width=8.5cm]{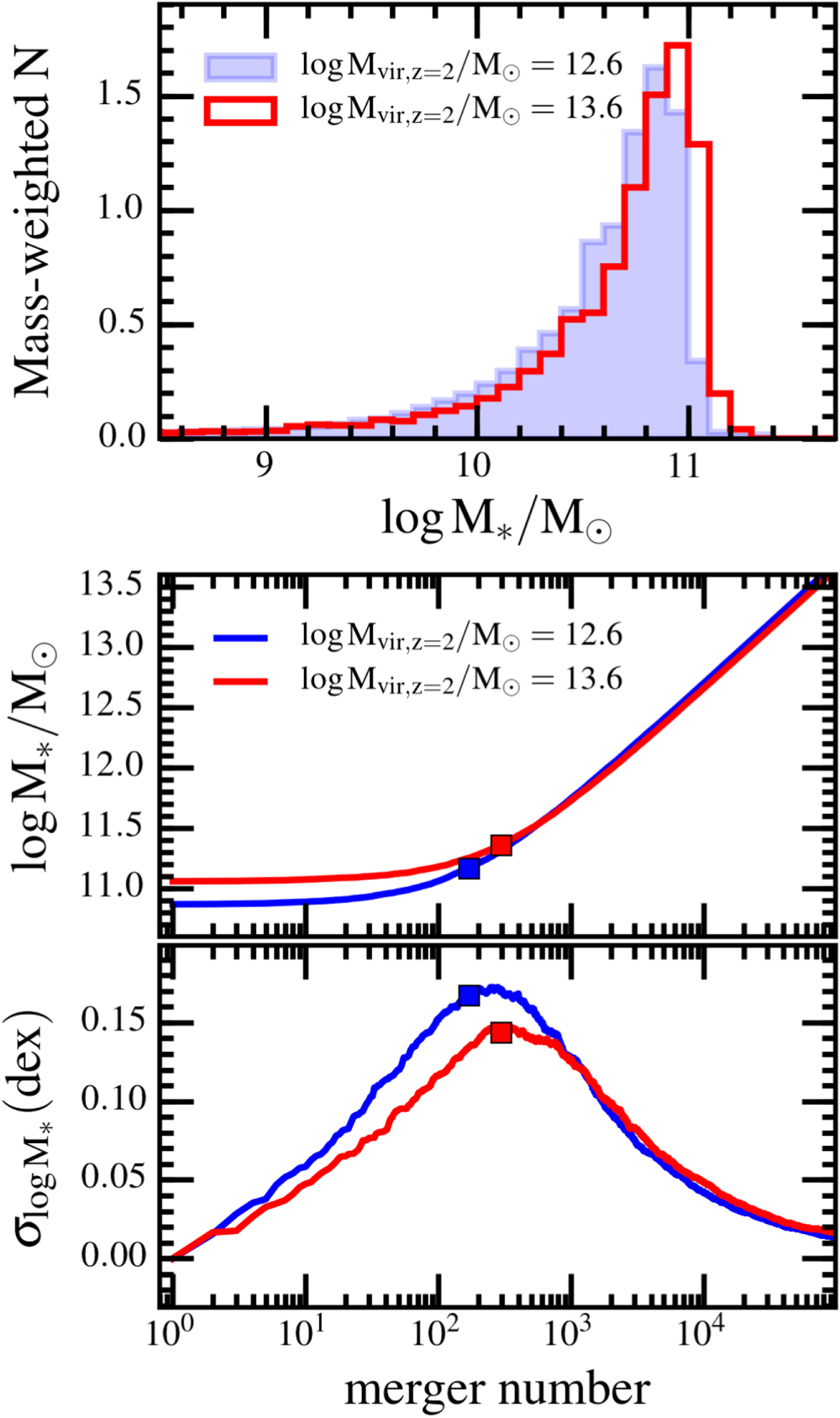}
\caption{
A simple Monte Carlo simulation for the stellar mass and scatter evolution due purely 
to hierarchical assembly in two halo mass bins.  Each halo merges with a random progenitor 
at each step.  Top panel shows the mass-weighted number distributions of progenitors 
of each halo mass bin.  These distributions are the actual ones of progenitor masses for the 
halos in each mass bin in our fiducial model.  Bottom panel shows the evolution of 
the average stellar mass and the scatter of 500 host halos as a function of 
the merger numbers. Squares indicate the point at which the 
mean stellar mass is doubled.  
\label{figure 7}}
\end{figure}

\subsection{In-situ Star Formation}

A generic expectation of modern galaxy formation models is that the 
mass buildup of galaxies is largely driven by internal (in-situ) processes at 
low masses and external (ex-situ) processes at high masses.  
In this section, we explore the response of the scatter in the SMHM relation 
to the combined effects of ex-situ and in-situ growth between $z=2$ and $0$. 
The in-situ model results in a final stellar mass that reproduces 
observational constraints \citep{Behroozi2013b} by construction.  We also include intrinsic 
scatter in the in-situ mode. 

The effect of including in-situ growth is shown in Figure~7.  Stellar mass 
growth due only to hierarchical assembly produces a final scatter of $0.16$~dex  
at $M_{\rm vir}>10^{14}{M_\odot}$ at $z=0$, and even larger at lower masses (red line).  
We then include the stellar mass growth due to in-situ processes between $z=2$ and $0$ 
following the procedure described in Section~2.4, and assume that both 
the in-situ and ex-situ mass growth have no intrinsic scatter ($\sigma_i = 0$). 
The inferred total in-situ fraction is shown in the top panel of Figure~7.  The final 
scatter of the SMHM relation is shown in blue. At the high mass end, it's very similar to 
our merger model, since the in-situ growth after $z=2$ for cluster scaled halos is small. 
This indicates that the scatter in the SMHM relation at the group and cluster scale is mostly 
determined by ex-situ growth.   

In addition to the diversity of merger histories, galaxies in dark matter halos of 
the same mass may experience different in-situ growth rates. One evidence 
of this is the well-defined relation between stellar mass and star 
formation rate, which has a scatter of $\sim0.3$~dex.  
Therefore at fixed halo mass we might expect a dispersion of stellar mass due to 
in-situ processes.  For this reason, we further consider a model in which the intrinsic 
scatter associated with the in-situ component has a value of $\sigma_i = 0.2$~dex.  
As described in Section~2.4, we assume that the total stellar masses due to 
in-situ growth at fixed halo mass at $z=0$ have a scatter of $\sigma_i = 0.2$~dex for host halos.   
The stellar masses of progenitors between $z=2$ and $0$ at fixed halo mass also have a 
scatter of $\sigma_i = 0.2$~dex.  In Figure~7, we plot the final scatter for host halos at $z=0$ for the 
$\sigma_i = 0.2$~dex case in black.  Since for low mass halos the mass buildup 
after $z=2$ is driven largely by in-situ growth, the scatter at the low mass end is mainly 
determined by the in-situ scatter we assumed ($\sigma_i = 0.2$~dex).  At the high mass end, 
the scatter is only slightly affected by the in-situ component.  The overall scatter is 
almost flat as a function of halo mass.  We show the result of the $\sigma_i = 0.3$~dex case in Figure~8. 

\subsection{Comparison to Empirical Estimates of the Scatter}
We now turn to a comparison with observations.  In Figure 8 we plot our 
model prediction for the scatter including the combined effect of in-situ and ex-situ stellar 
mass growth. We consider three values for the scatter due to in-situ processes: $\sigma_i=0$, 
$0.2$~dex and $0.3$~dex.  These three options are shown as 
dashed, solid and dotted lines. 

First, we compare our result with direct observations. \cite{Kravtsov2014} presented
measurements of the stellar mass and halo mass for 21 clusters using optical, infrared and
X-ray data, and concluded that the relation between the stellar mass of the brightest
cluster galaxies (BCGs) and $M_{500}$ has a scatter of $\sigma_{\log M_{\ast,BCG}} = 0.17
\pm 0.03$. \cite{Patel2015} provided the SMHM relation measured for low mass groups
between $0.5 \leq z \leq 1$.  They found that the observed scatter of this relation is
about $\sigma_{\log M_{\ast}} = 0.25$ dex.  Note that their observed scatter included the 
uncertainty from the stellar mass measurement.  The intrinsic scatter must be even smaller.

Some previous studies assumed that $\sigma_{\log M_{\ast}}$ is a constant for all halo 
mass.  \cite{Leauthaud2012} studied the evolution of the SMHM 
relation from $z=1$ to $0.2$ by analyzing the
galaxy-galaxy weak lensing, galaxy spatial clustering and galaxy number densities from the
COSMOS survey. They had two models for the scatter.  In the first model the $\sigma_{\log
M_{\ast}}$ was assumed to be constant.  In the second model, they assumed that the total 
scatter is the sum in quadrature of a constant intrinsic scatter and a mass-dependent 
scatter for the measurement error part.  They fitted for the intrinsic scatter in the second 
model.  They concluded that the two models produce very similar results.
In Figure~8 we compare our result with the intrinsic scatter from their second model:
$\sigma_{\log M_{\ast}} = 0.192 \pm 0.031$ dex at $0.22 \leq z \leq 0.48$ and
$\sigma_{\log M_{\ast}} = 0.220 \pm 0.019$ dex at $0.74 \leq z \leq 1$.  \cite{Yang2009}
constrained the relation between halo mass and stellar mass of central galaxies 
using the observed stellar mass function provided by the
galaxy group catalog from the SDSS DR4.  They concluded the stellar mass distribution of 
galaxies at fixed halo mass can be described by a log-normal distribution and the scatter 
is roughly $0.17$~dex. We compare our result with theirs from a combined sample of red and blue galaxies.

\begin{figure}[t] 
\includegraphics[width=8.5cm]{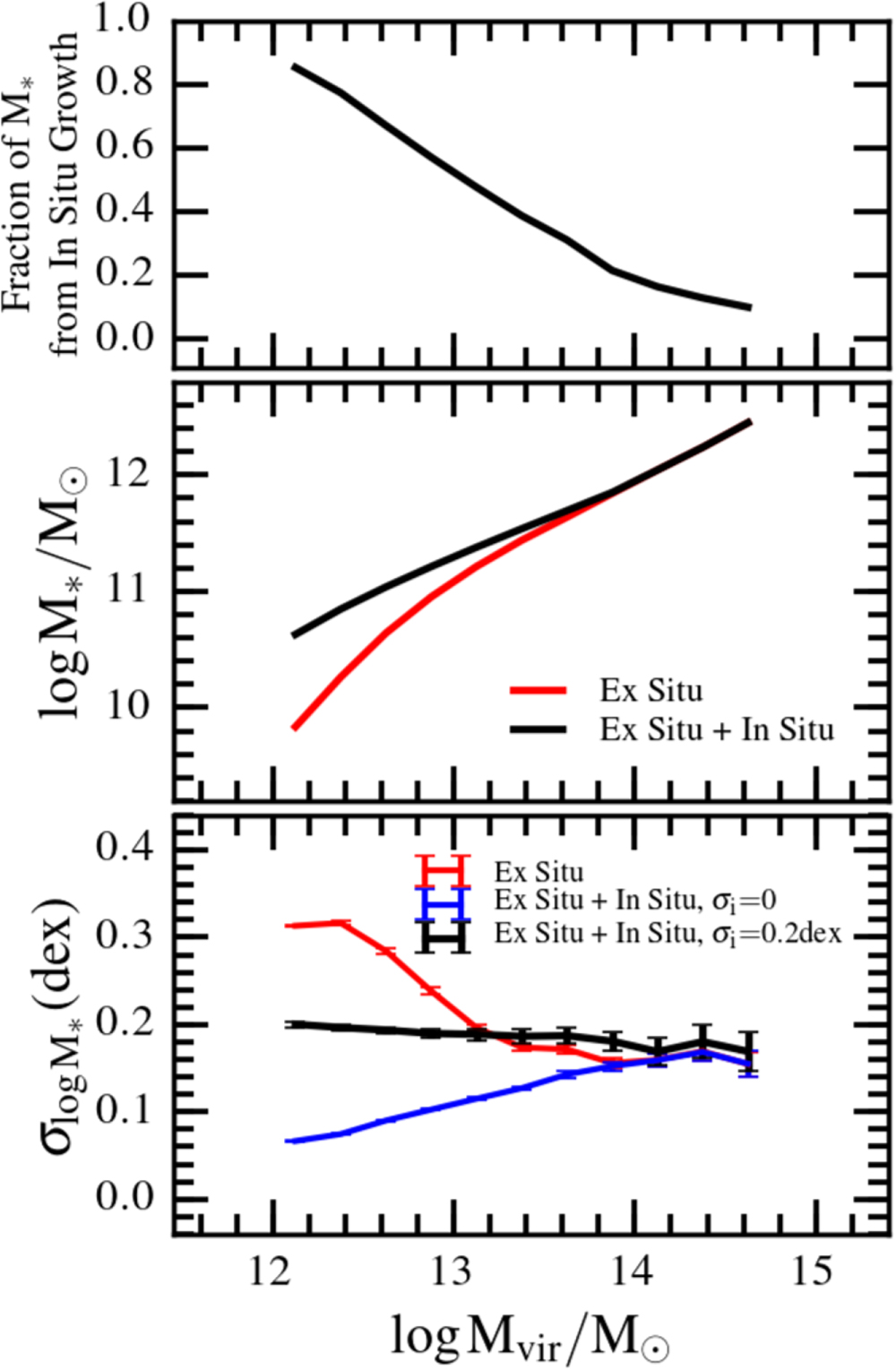}
\caption{
Top panel: The stellar mass fraction from in-situ growth as a function of halo mass 
at $z=0$. Middle panel:  The evolved SMHM relation at $z=0$ solely driven by the 
hierarchical assembly of dark matter halos since $z=2$(red), 
and by both ex-situ and in-situ processes (black).  Bottom:  The evolved scatter in 
$\log M_{\ast}$ at fixed halo mass at $z=0$ solely driven by ex-situ processes (red), by both mergers  
and in-situ growth with zero intrinsic scatter (blue), and by both mergers and an in-situ growth with 
a $0.2$~dex scatter (black).
\label{figure 3}}
\end{figure}

\noindent

\cite{More2011} used the kinematics of satellite galaxies and the SDSS data to study the
SMHM relation of central galaxies, and found the dependency of scatter on the color.  They 
found that the scatter of halo mass at fixed stellar mass is constant for blue galaxies but 
increases as a function of stellar mass for red galaxies.  Their results showed that
$\sigma_{\log M_{\ast}} = 0.17^{+0.04}_{-0.03}$~dex for red central galaxies and
$\sigma_{\log M_{\ast}} = 0.15^{+0.11}_{-0.08}$~dex for blue central galaxies. 
We only plot their result for red central galaxies in Figure~8 for simplicity.  
\cite{Tinker2013} also investigated the dependency on color at 3 redshift bins, $[0.22,0.48],
[0.48,0.74], [0.74,1]$. They constrained the SMHM relation for star-forming, and passive
galaxies separately by combining constraints from the stellar mass function, the 
angular correlation function and galaxy-galaxy lensing from COSMOS data.  
In the low redshift bin, $\sigma_{\log
M_{\ast}} = 0.21 \pm 0.06$~dex for star-forming galaxies and $\sigma_{\log M_{\ast}} =
0.28 \pm 0.03$~dex for passive galaxies.  In the high redshift bin, $\sigma_{\log
M_{\ast}} = 0.25 \pm 0.01$~dex for star-forming galaxies and 
$\sigma_{\log M_{\ast}} = 0.18 \pm 0.05$~dex for passive galaxies.  
In Figure~7 we compare our result with theirs for passive galaxies in their highest and lowest redshift bins. 

\cite{Reddick2013} used subhalo abundance matching as well as constraints from the
projected two-point galaxy clustering and the observed conditional stellar mass function
to model the relation between the stellar mass and the peak circular velocity.  They
concluded that the peak circular velocity of the halos is the property that most closely
connected to galaxy stellar mass, and the scatter in stellar mass at fixed peak velocity
is $\sigma_{\log M_{\ast}} = 0.20 \pm 0.03$ dex.  However, they also constrained the peak
velocity-dependent scatter by only using the conditional stellar mass function for
independent halo mass bins.  They concluded that the scatter measured for the independent
halo mass bins is consistent with their first model.  

\cite{Lehmann2015} also conducted a subhalo abundance matching analysis by matching 
galaxy luminosity to halo properties.  They included a new parameter, $\alpha$, to control the 
dependence on concentration when halos are ranked by mass.  They found a degeneracy between 
$\alpha$ and the log-normal scatter of galaxy luminosity.  Moreover, their results show that the 
most important constraint of scatter is the clustering at high luminosity end, while clustering 
at lower luminosity ($\sim L_*$) mainly constrains $\alpha$. They further assumed that both 
$\alpha$ and scatter are constant as a function of galaxy luminosity and found that best-fit 
value at $z=0.05$ is $\sigma_{\log L} = 0.17^{+0.03}_{-0.05}$~dex.

Comparison between the data and our models suggests that the scatter 
associated with in-situ growth should be limited to $\approx0.2$~dex 
in order to reproduce a flat scatter in the SMHM relation from 
$10^{12}{M_\odot}<M_{\rm vir}<10^{14.75}{M_\odot}$.  At the group and cluster scale 
($M_{\rm vir}>10^{14}{M_\odot}$), the scatter in the SMHM relation is not sensitive to 
in-situ process and appears to be a generic outcome of hierarchical assembly of massive galaxies.   

\section{Discussion}

We have interpreted the scatter in the SMHM relation as a result of two 
distinct processes.  At the high mass end the scatter is determined by hierarchical 
assembly, while at low masses the scatter is shaped by the intrinsic scatter associated 
with in-situ growth. Observations such as galaxy-galaxy lensing, satellite 
kinematics and galaxy clustering provide constraints on scatter for halos with 
$M_{\rm vir}>10^{12}{M_\odot}$. They suggest that scatter associated with in-situ 
growth is limited to $\approx0.2$~dex.  At high masses hierarchical growth also produces 
a scatter of $\approx 0.2$~dex in stellar mass at fixed halo mass. 

\begin{figure*}[htb] 
\includegraphics[width=17cm]{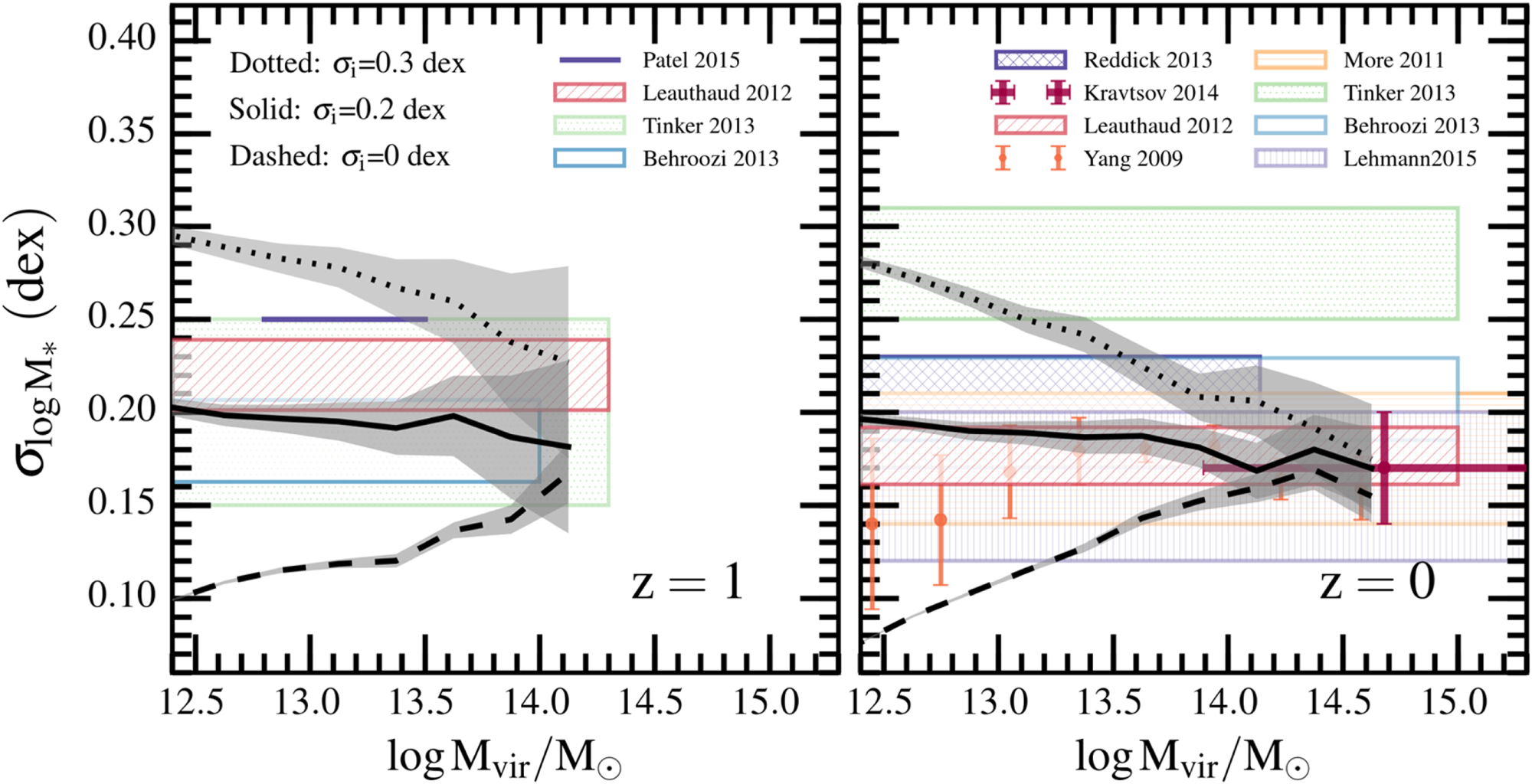}
\caption{
The scatter of stellar mass at fixed halo mass due to the combined effects from mergers and 
in-situ star formation.  Left and Right panels show the predicted scatter from our models 
at $z=1$ and $z=0$ respectively, 
and are compared to observational estimates.  Dotted, solid and dashed lines show the 
scatter assuming a $0.3$~dex, $0.2$~dex and zero dispersion in mergers and in-situ star formation, 
respectively. Shaded regions represent uncertainties of the scatter determined by 
bootstrapping.
\label{figure 8}}
\end{figure*}

Our fiducial ex-situ growth model indicates that at $z=0$, the scatter of the SMHM 
relation is strongly mass-dependent, reaching $0.32$~dex at $M_{\rm vir}\approx10^{12.5}{M_\odot}$ and 
$0.16$~dex at $M_{\rm vir}\approx10^{14.75}{M_\odot}$.  This result alone has interesting implications.  
We should expect galaxies between $10^{10}$ to $10^{11} M_{\odot}$ to have larger 
scatter in their ex-situ stellar mass fraction.  Assuming that most stars in the stellar 
halos come from ex-situ processes, our ex-situ only model provides a novel prediction for the scatter in the mass in the stellar halos.  Upcoming surveys (e.g., \citealp{vanDokkum2014}) of stellar halo will help to 
constrain this picture.

The scatter of the SMHM relation contains valuable information on the galaxy formation process.  
It deserves careful consideration in both observations and simulations in the future.  At the 
high mass end ($M_{\rm vir}>10^{14}{M_\odot}$), the scatter is a generic consequence of the 
hierarchical assembly.  Different simulations should be able to produce consistent scatter 
of the SMHM relation in this mass range.  At lower halo mass ($M_{\rm vir}<10^{13}{M_\odot}$) 
the scatter gradually becomes dominated by baryonic processes involved in the in-situ build-up, 
and has potentially more constraining power on the underlying theory.  

Two different processes, in-situ and ex-situ growth are needed to produce the apparent 
flat scatter in the SMHM relation.  However the similar magnitude of the observed scatter 
at the low mass end and high mass end is likely a coincidence. Other works (e.g., 
\citealp{BoylanKolchin2012, GarrisonKimmel2014, Sawala2015, Dutton2015}) have shown that at the 
regime of dwarf galaxies, the scatter of the SMHM relation may be much larger than in 
more massive systems.  More stringent observational constraints on the mass dependence of the 
scatter would be valuable.
 
We end this section by highlighting several caveats and areas for future improvement.  
The Bolshoi simulation has a limited resolution 
(one particle mass~$=1.9 \times 10^8 {M_\odot}$) and so halos with mass smaller than 
$2\times10^{10} {M_\odot}$ may not be properly described in the Bolshoi
simulation.  As shown in Figure~1, a fraction of satellite galaxies are possibly affected
by this limited resolution.   The result for host halo at the low mass end ($M_{\rm
vir, z=0} <10^{13} {M_\odot}$) should be treated with caution.  However as discussed earlier,
since the stellar component in such halos should be very small based on the SMHM relation,
this should only have a small effect on our result.   Moreover, we adopt a simple
assumption that when a subhalo disrupts, all of its stellar content is accreted to the
host halo, while in reality,  the accretion of stellar mass could be delayed
(\citealp{Wetzel2010}).  We also do not distinguish the stellar mass in the central galaxy
and the ICL.  Since most observations only include the stellar mass measurement of the
central part of the host halo, a direct comparison with observations should also be
treated with caution.  The scatter in the SMHM relations observed should be roughly the 
quadratic sum of the intrinsic scatter discussed above and the measurement error.  
\cite{Behroozi2010} carefully studied different uncertainties in the SMHM relation.  The 
authors included statistical uncertainties in the stellar mass function, cosmological 
parameters, and the uncertainty within the methodology to construct the SMHM relation, 
and revealed a $\sim 0.07$ dex (and $\sim 0.12$ dex) observational uncertainty at $z=0$ 
(and $z=1$).  As shown in Figure 8, we compare with the observed scatter at $z=0$ 
from \cite{Kravtsov2014}.  Even without the observational error, our fiducial model 
including a $0.2$~dex scatter due to intrinsic process is already comparable with the 
observation at $z=1$ and $z=0$.
If we consider the statistical errors provided by \cite{Behroozi2010}, the intrinsic 
scatter could be even smaller.

\section{Conclusions}

We have used the Bolshoi simulation to study the origin of the scatter 
in the SMHM relation.  We have included stellar mass growth both due to hierarchical 
assembly (ex-situ) and in-situ processes.  Our main results are summarized as follows. 

\begin{itemize}
  \item The scatter due to hierarchical assembly is mass-dependent.  At the group and cluster 
  scale ($M_{\rm vir}>10^{14}{M_\odot}$) the scatter due purely to mergers is 
  $\approx0.16$~dex, which is very close to recent observational estimates.  
  At lower halo masses ($M_{\rm vir}\sim10^{12}{M_\odot}$) the scatter 
  increases to $0.32$~dex.  Stellar halos, which are likely the result of this hierarchical 
  assembly, may therefore show factors of $\approx2$ scatter from galaxy to galaxy at this mass range.
  (Section~3.1.1)
  \item Several factors influence scatter growth.
  The scatter is affected by the average number of mergers a population has experienced.  The 
  slope of the SMHM relation affects the scatter in a way that a steeper relations produces a 
  larger scatter.  The growth of dark matter halos due to smooth accretion has a 
  significant effect on the scatter growth.
  (Section~3.1.2-3.1.4)
  \item At lower masses, mass buildup since $z\approx2$ is driven largely by in-situ growth.  We 
  include a model for the in-situ buildup of stellar mass.  We find that an intrinsic scatter in 
  this growth channel of $0.2$~dex produces a relation between scatter and halo mass that is 
  consistent with observations from $10^{12}{M_\odot}<M_{\rm vir}<10^{14.75}{M_\odot}$.
  (Section~3.2-3.3)
\end{itemize} 

The scatter in the SMHM relation is affected by two distinct processes: hierarchical assembly 
and in-situ growth. More precise observational estimates of the mass-dependent scatter should 
constrain both of these effects, and in particular should provide 
new insights into the regularity of the galaxy formation process.

\acknowledgements
We are grateful to Andrey Kravtsov and Risa Wechsler for helpful comments on 
an early draft.  M. G. acknowledges support from the National Science Foundation 
Graduate Research Fellowship.  C. C. acknowledges support from the Packard 
Foundation.  P. B. was partially supported by a Giacconi Fellowship from the 
Space Telescope Science Institute. The remainder of support for P. B. through 
program number HST-HF2-51353.001-A was provided 
by NASA through a Hubble Fellowship grant from the Space Telescope Science Institute, 
which is operated by the Association of Universities for Research in Astronomy, 
Incorporated, under NASA contract NAS5-26555.  The computations in this paper 
were run on the Odyssey cluster supported by the FAS Division of Science, Research Computing Group at Harvard University.  This work made 
use of the Bolshoi simulation.  The Bolshoi simulation has been performed within the Bolshoi 
project of the University of California High-Performance AstroComputing Center (UC-HiPACC) 
and were run at the NASA Ames Research Center.

\end{document}